\documentclass{wlscirep}
\usepackage[utf8]{inputenc}
\usepackage[T1]{fontenc}
\usepackage{times}
\usepackage{graphicx}
\usepackage{amssymb,amsmath}
\usepackage{xcolor}
\usepackage{soul}
\usepackage{lineno}
\usepackage{caption}
\DeclareCaptionLabelSeparator{bar}{ \textbar\ }
\newcommand\apj{Astrophys. J.}
\newcommand\apjl{Astrophys. J. Lett.}     
\newcommand\apjs{Astrophys. J. Suppl.}
\newcommand\aap{Astron. Astrophys.}
\newcommand\aapr{Astron. Astrophys. Rev.}
\newcommand\mnras{Mon. Not. R. Astron. Soc.}
\newcommand\pasp{Publ. Astron. Soc. Pacific}
\newcommand\solphys{Sol. Phys.}
\newcommand\ssr{Space Sci. Rev.}
\newcommand\nat{Nature}
\newcommand\sci{Science}
\newcommand\jgr{J. Geophys. Res.}
\newcommand\lrsp{Liv. Rev. Sol. Phys.}
\newcommand\natastro{Nat. Astron.}
\newcommand\jpp{J. Plasma Phys.}
\newcommand\pplasma{Phys. Plasmas}
\newcommand\rnaas{Res. Notes Am. Astron. Soc.}
\newcommand{\edit}{\textcolor{black}}
 
\title{Measurement of magnetic field and relativistic electrons along a solar flare current sheet} 

\author[1]{Bin Chen\footnote{Correspondence to \href{bin.chen@njit.edu}{bin.chen@njit.edu}}}
\author[2]{Chengcai Shen}
\author[1]{Dale E. Gary}
\author[2]{Katharine K. Reeves}
\author[1]{Gregory D. Fleishman}
\author[1]{Sijie Yu}
\author[3,4]{Fan Guo}
\author[5,6]{S\"am Krucker}
\author[7,8,9]{Jun Lin}
\author[1]{Gelu Nita}
\author[10]{Xiangliang Kong}
\affil[1]{Center for Solar-Terrestrial Research, New Jersey Institute of Technology, Newark, NJ 07102, USA}
\affil[2]{Harvard-Smithsonian Center for Astrophysics, Cambridge, MA 02138, USA}
\affil[3]{Los Alamos National Laboratory, Los Alamos, NM 87545, USA}
\affil[4]{New Mexico Consortium, Los Alamos, NM 87544, USA}
\affil[5]{University of California, Berkeley, Berkeley, CA 94720, USA}
\affil[6]{University of Applied Sciences and Arts Northwestern Switzerland, 5210 Windisch, Switzerland}
\affil[7]{Yunnan Observatories, Chinese Academy of Sciences, Kunming, Yunnan 650216, China}
\affil[8]{Center for Astronomical Mega-Science, Chinese Academy of Sciences, Beijing 100012, China}
\affil[9]{University of Chinese Academy of Sciences, Beijing 100049, China}
\affil[10]{Institute of Space Sciences, Shandong University, Weihai, Shandong 264209, China}

\begin{abstract}
\textbf{In the standard model of solar flares, a large-scale reconnection current sheet is postulated as the central engine for powering the flare energy release \cite{2000JGR...105.2375L,1994Natur.371..495M,2011LRSP....8....6S} and accelerating particles \cite{1996ApJ...462..997L,2006Natur.443..553D,2011ApJ...737...24B}.
However, where and how the energy release and particle acceleration occur remain unclear due to the lack of measurements for the magnetic properties of the current sheet. Here we report the measurement of spatially-resolved magnetic field and flare-accelerated relativistic electrons along a current-sheet feature in a solar flare. The measured magnetic field profile shows a local maximum where the reconnecting field lines of opposite polarities closely approach each other, known as the reconnection $X$ point. The measurements also reveal a local minimum near the bottom of the current sheet above the flare loop-top, referred to as a ``magnetic bottle''. This spatial structure agrees with theoretical predictions \cite{2000JGR...105.2375L,2018ApJ...858...70F} and numerical modeling results. A strong reconnection electric field of $\sim$4000 V m$^{-1}$ is inferred near the $X$ point. This location, however, shows a local depletion of microwave-emitting relativistic electrons. \edit{These electrons concentrate instead} at or near the magnetic bottle structure, where more than 99\% of them reside at each instant. Our observations suggest \edit{that the loop-top magnetic bottle is likely the primary site for accelerating and/or confining the relativistic electrons}.}
\end{abstract}

\begin{document} 
\flushbottom
\maketitle 

Our measurement of the magnetic field and the relativistic electrons is made possible by microwave spectral imaging observations of a large X8.2 solar flare on 2017 September 10 (the second largest in Solar Cycle 24) from the newly commissioned Expanded Owens Valley Solar Array (EOVSA) \cite{2018ApJ...863...83G}. \edit{This event is associated with a fast coronal mass ejection (CME) that drives a large-scale coronal shock in the upper solar corona \cite{2019NatAs...3..452M}. During the initial phase of the event, the CME is observed in the low corona as a rapidly ascending, balloon-shaped dark cavity at extreme ultraviolet (EUV) and microwave wavelengths (Fig. \ref{fig:overview}), interpreted as an erupting magnetic flux rope viewed along its axis} \cite{2018ApJ...853L..18Y,2018ApJ...868..107V, 2020arXiv200501900C}. This flux rope is connected to the top of newly reconnected, cusp-shaped flare arcade by a thin elongated plasma sheet, presumably associated with a large-scale reconnection current sheet (RCS), extending down from the bottom of the cavity. The plasma sheet appears bright in EUV bands sensitive to hot flare plasma (Fig. \ref{fig:spec}(a)) but dark in EUV bands sensitive to background coronal temperatures (Fig. \ref{fig:spec}(b)), indicating that it has undergone intense flare heating \cite{2018ApJ...853L..18Y,2018ApJ...854..122W, 2018ApJ...868..148L}. Despite the slight asymmetry of the cusp-shaped flare arcade (Fig. \ref{fig:spec}(a)), the observed features in the plane of the sky offer an ideal case to test against the theoretical predictions. Indeed, thanks to the favorable viewing perspective, these features match very well the overall magnetic configuration in one of most well-known theoretical standard flare models by Lin \& Forbes\cite{2000JGR...105.2375L} depicted in two dimensions (white curves in Figs. \ref{fig:overview}(a) and (b); Methods). 

\renewcommand\thefigure{\arabic{figure}}
\renewcommand{\figurename}{Fig.}
\setcounter{figure}{0} 
\begin{figure*}[!ht]
\begin{center}
\includegraphics[width=0.95\textwidth]{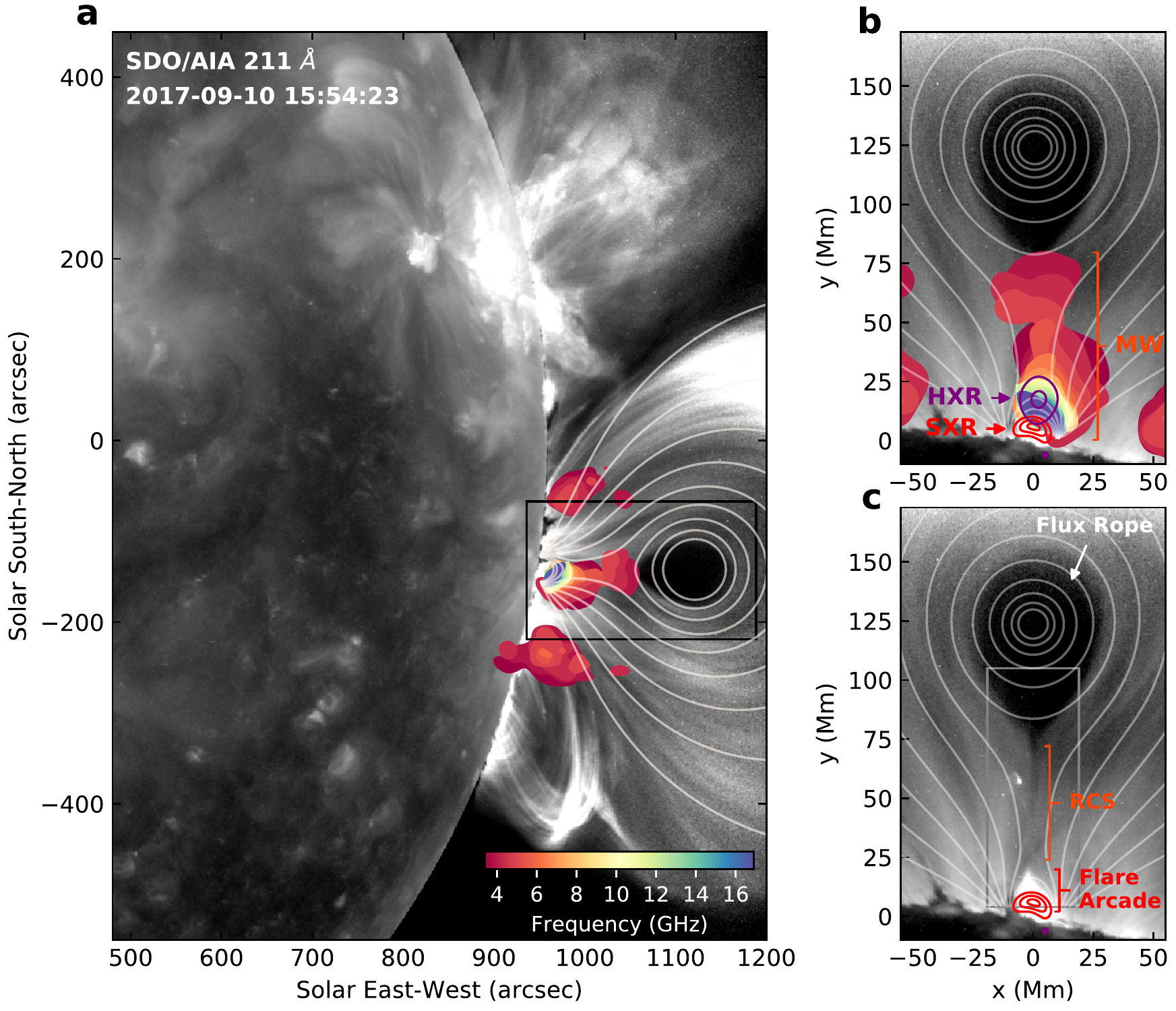}
\end{center}
\caption{\textbf{Observation and modeling of the eruptive solar flare on 2017 September 10.} (\textbf{a}) EUV 211 \AA\ image showing the erupting magnetic flux rope as a fast-ascending balloon-shaped dark cavity, observed by the Atmospheric Imaging Assembly aboard the Solar Dynamics Observatory (SDO/AIA). The multi-frequency EOVSA microwave (MW) source is shown as filled color contours (26\% of the maximum intensity at their respective frequency). White curves are magnetic field lines derived from the theoretical standard flare model in Lin \& Forbes\cite{2000JGR...105.2375L}. (\textbf{b}) Detailed view of the central region (black box in \textbf{a}, rotated by 90$^{\circ}$ to upright orientation). A 30--100 keV hard X-ray (HXR) source (purple contours; showing 50\% and 90\% of the maximum), observed by the Reuven Ramaty High Energy Solar Spectroscopic Imager (RHESSI), is present above the top of the soft-X-ray-emitting hot flare arcade (red contours; showing 30\%, 60\%, and 90\% of the maximum intensity at 12--18 keV, also observed by RHESSI). (\textbf{c}) Same field of view as (b), but the nearly identical magnetic field lines are derived from the numerical magnetohydrodynamics simulation (see Methods). The microwave and HXR source are removed to show the cusp-shaped EUV flare arcade (bright white).}\label{fig:overview}
\end{figure*}

EOVSA microwave spectral imaging observations provide a picture of the flare-accelerated electrons with energies extending to at least hundreds of keV in the relativistic regime \cite{2018ApJ...863...83G}. During the primary flux rope acceleration and energy release phase around 15:54 UT \cite{2018ApJ...868..107V}, the microwave-emitting relativistic electrons are present throughout the entire region between the erupting flux rope and the flare arcade where the RCS is located (filled contours in Fig. \ref{fig:overview}(b)). The multi-frequency microwave source resembles an ``hourglass'' shape: The upper part starts from the bottom of the flux rope and narrows downward, then joins its lower counterpart located above the flare arcade that broadens toward lower heights.

\begin{figure*}[!ht]
\begin{center}
\includegraphics[width=1.0\textwidth]{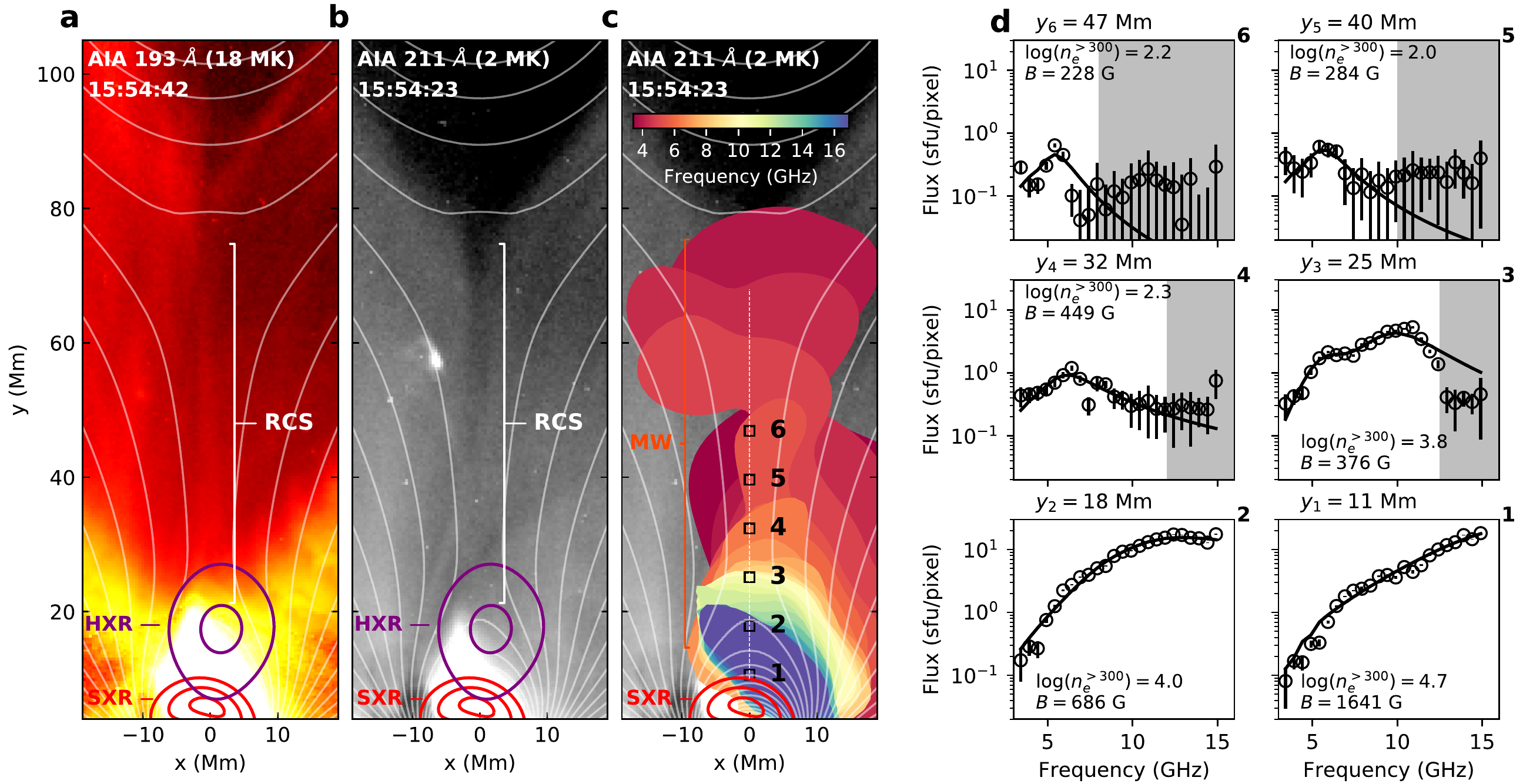}
\end{center}
\caption{\textbf{Spatially-resolved microwave spectra in the reconnection current sheet region.} In the enlarged view of the central region (gray box in Figure \ref{fig:overview}(c)), the RCS can be identified as a thin elongated feature near $x=0$ Mm, which appears bright in SDO/AIA EUV 193 \AA\ band sensitive to heated plasma of $\sim$18 MK \textbf{(a)}, and dark in EUV 211 \AA\ band sensitive to cooler coronal plasma of $\sim$2 MK \textbf{(b)}. (\textbf{c}) Same as (b), but with the multi-frequency microwave source overlaid. (\textbf{d}) Examples of the microwave spectra (circle symbols with error bars) from selected locations along the RCS feature at $x\approx 0$ Mm (numbered small boxes in (c)). The error bars show the uncertainties evaluated by using the root-mean-square of the background fluctuations in an area away from the source. Shaded areas indicate dynamic-range-limited data points excluded from the spectral fit. The corresponding best-fit results based on gyrosynchrotron radiation are shown as black curves. Also shown are the corresponding values of the magnetic field strength ($B$, in Gauss) and relativistic electron density with energy above 300 keV ($n^{>300}_e$, in cm$^{-3}$) from the spectral fit results.}\label{fig:spec}
\end{figure*}

\begin{figure*}[!ht]
\begin{center}
\includegraphics[width=1.0\textwidth]{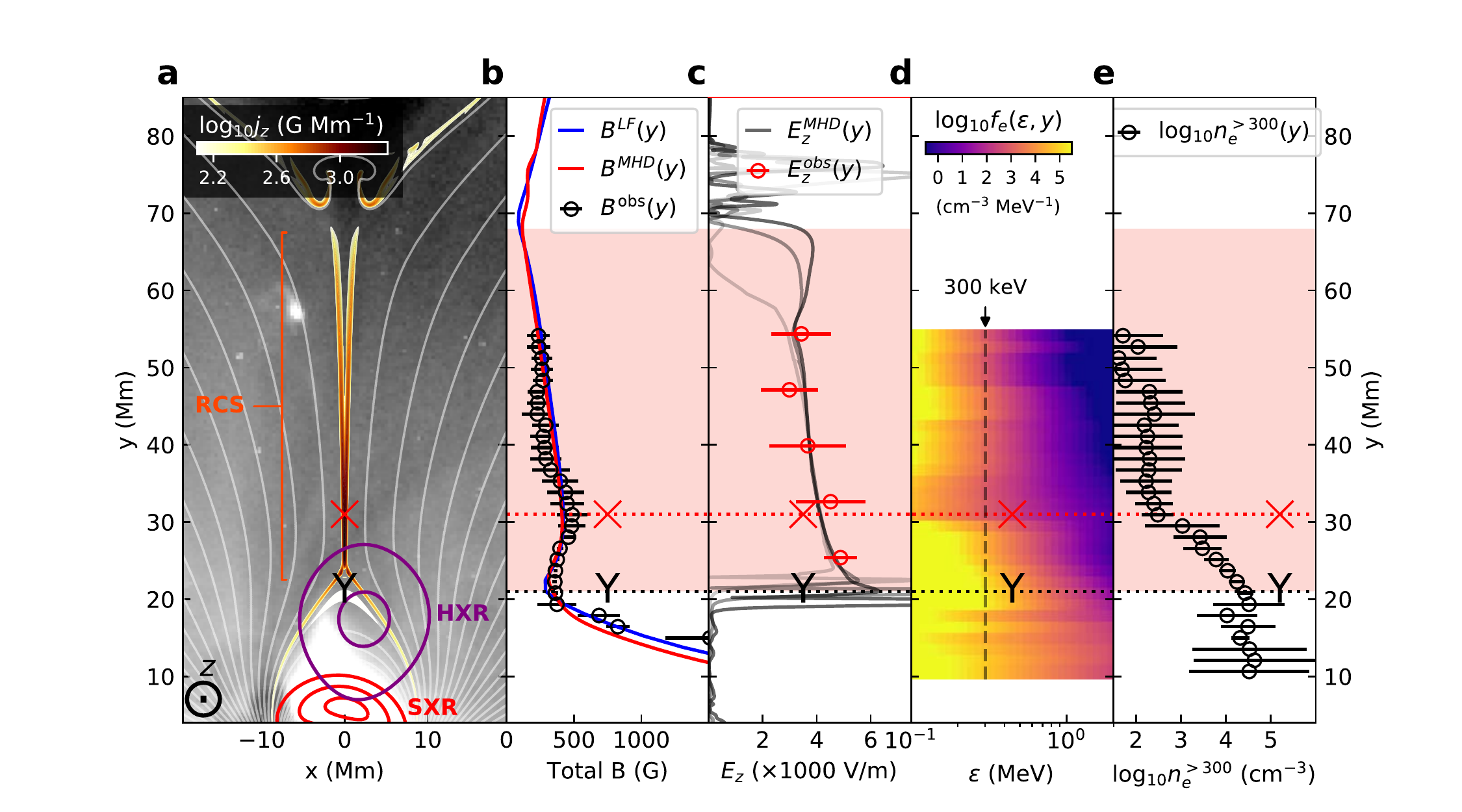}
\end{center}
\caption{\textbf{Spatial distribution of current density, magnetic field, electric field, and relativistic electrons along the reconnection current sheet.} (\textbf{a}) Similar to Fig. \ref{fig:spec}(b), to which has been added the electric current density distribution $j_z$ ($z$ is the direction perpendicular to the $x$-$y$ plane) derived from MHD simulation. (\textbf{b}) Measurements of the height profile of the total magnetic field strength along the RCS at $x\approx 0$ Mm ($B^{\rm obs}(y)$; black symbols), which agree with predictions of the theoretical standard flare model in Lin \& Forbes \cite{2000JGR...105.2375L} ($B^{\rm LF}(y)$; blue curve) and MHD simulation ($B^{\rm MHD}(y)$; red curve) also obtained at $x = 0$ Mm (after convolution with instrument resolution; Methods). (\textbf{c}) Distribution of the reconnection electric field along the RCS as a function of height estimated from the observations (i.e., $E^{\rm obs}_{z}(y)$, the electric field component perpendicular to the $x$-$y$ plane; red symbols). Light to dark red curves show the electric field obtained from the MHD simulation $E^{\rm MHD}_{z}(y)$ at selected locations close to the RCS (at $x=1, 2, 3$ Mm), multiplied by a factor of 3.5. (\textbf{d}) Height--energy diagram of the spatially resolved energetic electron energy distribution along the RCS derived from the microwave data ($f_e(\varepsilon,y)$). Color scale of the diagram represents the logarithm of the electron number density differentiated in energy. The corresponding spectral index of the electron energy distribution in the RCS region $\delta\approx 3$--6.
(\textbf{e}) Variation of relativistic electron density above 300 keV along the RCS ($n^{>300}_e(y)$). Horizontal bars on all values shown in panels (b), (c), and (e) represent the estimated uncertainties of the corresponding parameters. The inferred locations of the reconnection $X$ and $Y$ point are marked as a red $X$ symbol and a black $Y$ symbol, respectively. Pink-shaded region indicates the height range where the RCS is present.} \label{fig:rec_e}
\end{figure*}

From any pixel of EOVSA's multi-frequency microwave images at a given time, a spatially-resolved microwave spectrum can be obtained (Fig. \ref{fig:spec}). The microwave spectra display features characteristic of gyrosynchrotron radiation produced by flare-accelerated energetic electrons gyrating in the flare magnetic field \cite{2018ApJ...863...83G}. By fitting each microwave spectrum with a gyrosynchrotron source model at a given spatial location along the RCS feature (at $x\approx 0$ Mm), we derive the spatially-resolved total magnetic field strength $B^{\rm obs}(y)$ and microwave-emitting energetic electron distribution $f_e(\varepsilon,y)=dn_e(\varepsilon,y)/d\varepsilon$ at different heights $y$ along the RCS (where $n_e$ is the energetic electron number density and $\varepsilon$ is the electron energy) (Methods; see Fig. \ref{fig:spec}(d) for examples). The resulting $B^{\rm obs}(y)$ profile, shown in Fig. \ref{fig:rec_e}(b), represents the height variation of the magnetic field strength measured over our resolution element ($\sim$3 Mm at 15 GHz) at the location of the RCS (Methods). It displays a general decrease of magnetic field strength in height, which meets the expectation that the source of the magnetic flux is rooted at the photosphere and opens up in the coronal volume. 

Intriguingly, this $B^{\rm obs}(y)$ profile shows a local maximum located close to the point where the hourglass-shaped upper and lower microwave source join together (at $y\approx 31$ Mm). In addition, a local minimum is present near the tip of the cusp-shaped EUV flare arcade (at $y\approx 21$ Mm). By comparing with the magnetic field profile derived from the analytical standard flare model in Lin \& Forbes \cite{2000JGR...105.2375L} (at $x=0$ Mm, after convolution with EOVSA's instrument resolution; blue curve in Fig. \ref{fig:rec_e}(b) denoted as $B^{\rm LF}(y)$), we conclude that said features in the measured magnetic field profiles match well the features unique to the large-scale RCS: the local maximum corresponds to the ``pinch point'', or ``$X$'' point, where the reconnecting magnetic field  external to the RCS are brought in by plasma inflows and bow inward. The local minimum is associated with the bottom of the RCS connecting to the tip of the cusp-shaped flare arcade, sometimes referred to as the $Y$ point \cite{priest_forbes_2000}. These measured magnetic properties place a firm verification for the presence of the RCS at the location where an apparent plasma sheet also appears in EUV images.

To investigate the plasma dynamics and energetics (which the analytical model does not provide), we perform a self-consistent magnetohydrodynamics (MHD) numerical simulation based on initial conditions similar to those in the analytical standard flare model and observational constraints (Methods). Our MHD simulation yields excellent agreement with the flare morphology and dynamics (Figure \ref{fig:overview}(c) and Extended Data Fig. \ref{sfig:model}). Further, the RCS is clearly seen in the MHD simulation as a thin and elongated feature with a strong electric current density $j_z$ at the same location as the EUV plasma sheet (Fig. \ref{fig:rec_e}(a)). The vertical component of the magnetic field vector $B_y$ quickly switches its sign across the current sheet, indicating ongoing magnetic reconnection (Extended Data Fig. \ref{sfig:mhd_cs}; Methods). Similar to the analytical model, the total magnetic field strength profile along the RCS $B^{\rm MHD}(y)$ achieves excellent agreement with the measurements (red curve in Figure \ref{fig:rec_e}(b); after convolution with instrument resolution, see Methods). Moreover, our MHD simulation explicitly pinpoints the site from which bi-directional reconnection outflows are ejected along the RCS (i.e., where the vertical component of the plasma speed $v_y=0$). This site, sometimes referred to as the ``stagnation point'', is located close to the reconnection $X$ point identified from the magnetic field profile---another feature predicted by the theoretical standard flare model \cite{2018ApJ...858...70F}.

\begin{figure*}[!ht]
\begin{center}
\includegraphics[width=0.9\textwidth]{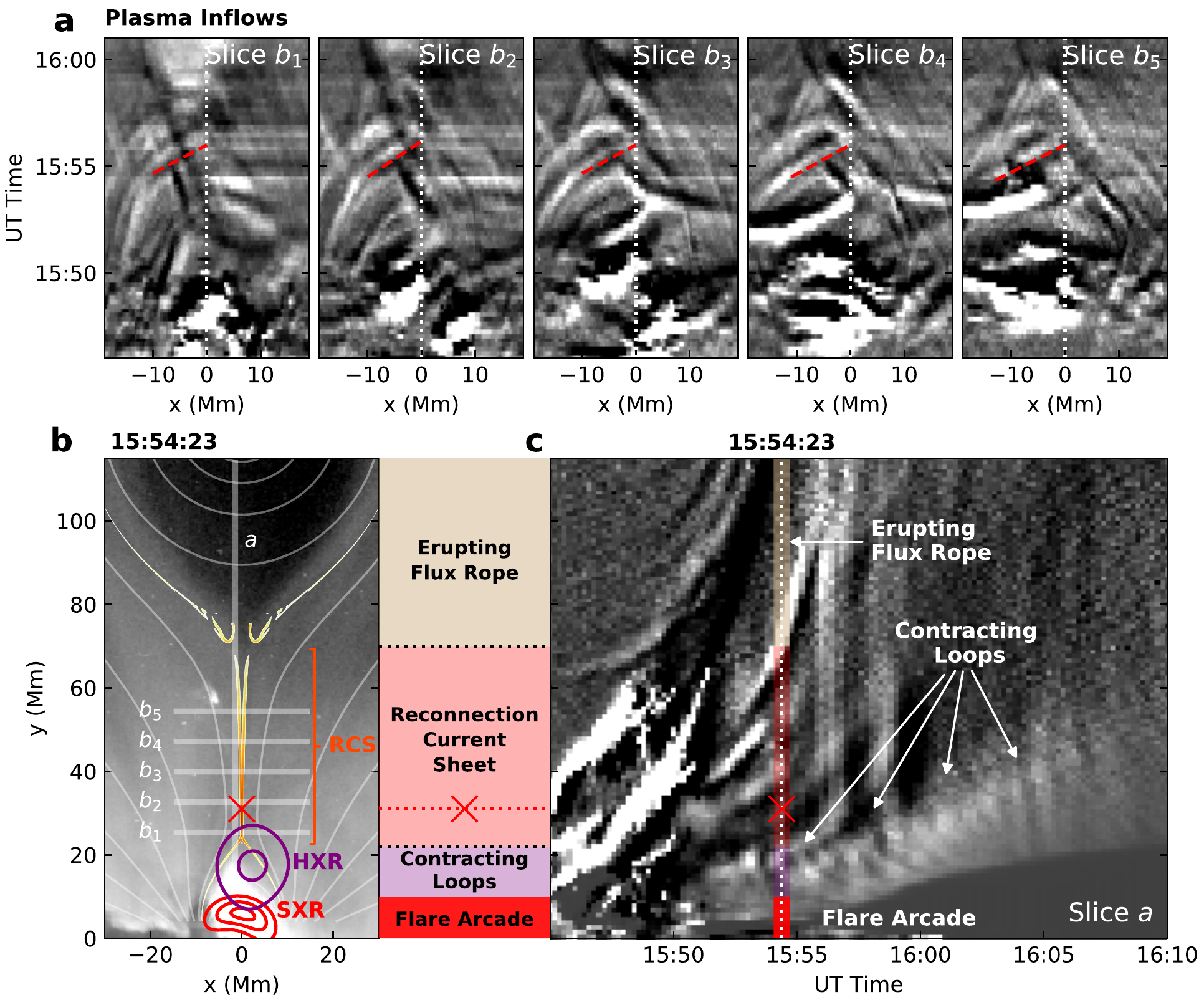}
\end{center}
\caption{\textbf{Plasma flows in the magnetic reconnection current sheet region.} (\textbf{a}) Time-distance diagrams showing plasma inflows toward the RCS observed in SDO/AIA EUV 171 \AA\ at different heights. Examples of the inflows are marked by red dashed lines, which have an average speed of $\sim$120 km s$^{-1}$. The corresponding horizontal slices for obtaining the diagrams are shown in (b) labeled from $b_1$ to $b_5$. (\textbf{b}) SDO/AIA EUV 211 \AA\ image and the corresponding MHD model (same as Fig. \ref{fig:rec_e}(a)). The ``$X$'' symbol indicating the location of the reconnection $X$ point. (\textbf{c}) Time-distance diagram obtained at a vertical slice ``a'' shown in panel (b). The vertical dotted line indicates the time of panel (b) at 15:54:23 UT. The upward erupting magnetic flux rope and downward contracting, newly reconnected magnetic loops are marked with arrows. An animation version of this figure is available as Supplementary Video 1.}\label{fig:flows}
\end{figure*}

EUV time-series imaging data provide means for directly measuring the speeds of inflowing plasma into the RCS (known as ``reconnection inflows'') at multiple heights $v_x(y)$ (Fig. \ref{fig:flows}(a); Methods), which are of order 100 km s$^{-1}$ throughout the RCS region (see also ref\cite{2018ApJ...853L..18Y}). The simultaneous and co-spatial measurements of $B$ and $v_x$ enable \edit{an estimate} for the spatial distribution of the electric field $E_z \approx v_x B_y/c$ and the electromagnetic energy (Poynting) flux $S_x \approx v_x B_y^2 /4\pi$ at the RCS. Here $B_y\approx B\sin\theta$ is the vertical component of the magnetic field strength in the close vicinity of the RCS.
$\theta$ is the viewing angle between the line of sight (LOS) direction $z$ and the magnetic field vector. It is a parameter constrained in our microwave spectral fitting, which is within 40--90$^\circ$ but has relatively large uncertainties (Methods). For the purpose of order-of-magnitude estimate, here we take the upper limit $B_y \approx B$, hence $E_z \approx v_x B/c$ and $S_x \approx v_x B^2 /4\pi$. Our estimate of the electric field in the RCS is over 4000 V m$^{-1}$ (red symbols in Fig. \ref{fig:rec_e}(c)), consistent with earlier indirect estimates \cite{2002ApJ...565.1335Q}. Such a strong electric field falls well into the super-Dreicer regime \cite{2002ApJ...565.1335Q}, which can easily accelerate electrons to relativistic energies (100s of keV to MeV) within a small acceleration distance of $\lesssim$1 km. The inflowing energy flux ${S_x}$  available for reconnection is of order $10^{10}$--$10^{11}$ ergs s$^{-1}$ cm$^{-2}$, sufficient to power a large X-class flare that releases $>$10$^{32}$ ergs in several minutes at its peak rate (Methods). The dimensionless reconnection rate $M = v_x/v_A$ is of order 0.01, where $v_A = 2\times10^{11} B/\sqrt{n_e^{\rm th}}\approx~$6,000--10,000 km s$^{-1}$ is the estimated Alfv\'en speed around the $X$ point with $B\approx300$--500 G (c.f., Fig. \ref{fig:rec_e}(b)) and thermal plasma density $n_e^{\rm th}$ of order $10^{10}$ cm$^{-3}$ (see, e.g., refs\cite{2018ApJ...854..122W,2018ApJ...868..148L} for estimates for $n_e^{\rm th}$ of this RCS feature \edit{and ref\cite{2008A&A...491..297R} for discussions on Alfv\'en speeds in solar active regions}).

We also derive the spatially-resolved energetic electron distribution along the RCS $f_e(\varepsilon,y)$ from the microwave data.  Fig. \ref{fig:rec_e}(d) shows this distribution as an energy--height diagram. In Fig. \ref{fig:rec_e}(e), we show the spatial distribution of the total electron number density at relativistic energies (integrated above 300 keV, or Lorentz factor of $>$1.6; i.e., $n_e^{>300}(y) = \int_{>300\ {\rm keV}}f_e(\varepsilon,y) d\varepsilon$). The microwave-emitting energetic electrons are ubiquitous throughout the RCS region. However, the shape of the spatial distribution of the relativistic electrons along the RCS $n_e^{>300}(y)$ does not demonstrate any obvious correlation with the reconnection electric field distribution $E_z(y)$ shown in Fig. \ref{fig:rec_e}(c). In particular, in the vicinity of the reconnection $X$ point (at $x\approx 31$ Mm), there exists a local depletion of the energetic electrons while a relatively hard electron energy spectrum is present (with a spectral index $\delta\approx3.3$, corresponding to a small color gradient over electron energy in Fig.~\ref{fig:rec_e}(d)). The hard spectrum suggests that the $X$ point might be a site for electron acceleration thanks to the presence of a strong electric field.
However, the relatively small number density of the energetic electrons indicates that either the acceleration efficiency is low around the $X$ point, or the electrons accelerated there escape rapidly and could not accumulate to an appreciable density \cite{2009JPlPh..75..159Z}. Such a depletion of energetic electrons, whether due to lack of acceleration or fast escape, may explain why microwave or HXR emission is often very weak or even entirely absent at the inferred reconnection $X$ point \cite{2003ApJ...596L.251S,2008ApJ...676..704L,2014ApJ...787..125N}.

In contrast, Fig. \ref{fig:rec_e}(d) shows that the spatial distribution of the energetic electrons $f_e$ at almost all energies strongly peaks in the vicinity of the $Y$ point near the bottom of the RCS, whereas the total number density of the relativistic electrons $n_e^{>300}$ exceeds those near the $X$ point by more than two orders of magnitude (Fig.~\ref{fig:rec_e}(e)). Thus, this region, which contains most of the microwave-emitting relativistic electrons, appears to be the primary site for confining and/or accelerating electrons to relativistic energies. It is also the site where HXR-emitting electrons at relatively low nonthermal energies (tens of keV) are frequently observed (purple contours in Fig. \ref{fig:rec_e}(a); see a review by ref\cite{2008A&ARv..16..155K}). This region coincides with the location where newly reconnected magnetic field lines emanating from the RCS interact vigorously with the underlying flare arcades, some of which are observed in EUV time-series images as multitudes of contracting loops (Fig. \ref{fig:flows}(c) and the accompanying animation; Methods). It has been proposed as a natural location for betatron acceleration by collapsing magnetic traps \cite{1997ApJ...485..859S} or Fermi-type acceleration processes that involve rapid contraction of magnetic islands \cite{2006Natur.443..553D} or plasma compression \cite{2018ApJ...866....4L}. Additionally, it provides an ideal environment for the generation of turbulence, waves, and (fractal) electric field \cite{2008ApJ...676..704L,2015ApJ...815....6Z,2018ApJ...854..122W} (see also a recent study by ref\cite{Fleishman2020} in which their presence is implied by an observed rapid decay of magnetic field), or ``termination'' shocks (formed by reconnection outflows impinging upon the flare arcade) \cite{1994Natur.371..495M,2015Sci...350.1238C,Kong_2019}, all of which have been suggested as possible particle acceleration mechanisms \cite{2011SSRv..159..357Z}. In addition to the plethora of likely acceleration processes, the local minimum of the magnetic field in this region represents a ``magnetic bottle'' to help confine electrons. Similar magnetic bottle structures have been observed \textit{in situ} in Earth's magnetosphere, within which an enhanced flux of energetic electrons and ions has been reported \cite{2019JGRA..124..197N}. 
The new methodology based on the microwave imaging spectroscopy reported here now permits the remote probing of such crucial plasma structures as solar flare RCSs.
These new measurements, representing 2D projections of three-dimensional (3D) physical phenomena in the plane of the sky, offer stringent constraints to guide theories of particle acceleration and advance realistic 3D modeling of solar flares.

\section*{Methods}
\subsection*{Magnetic Modeling}
Magnetic modeling of this event is performed along two lines, one based on a well-developed analytical model, and another based on a self-consistent, two-and-half-dimensional (2.5-D) resistive magnetohydrodynamic (MHD) numerical simulation, detailed below. 

\smallskip

\noindent \textbf{Analytical Model} \\
First, we investigate the general geometry of the event and magnetic field profile by adopting a analytical eruptive flare model first developed by Priest \& Forbes \cite{1990SoPh..126..319P, 1995ApJ...446..377F}, which was then further refined by several works including refs\cite{2000JGR...105.2375L,  2005ApJ...630.1133R, 2018ApJ...858...70F}. This model, sometimes referred to as the ``catastrophe model'', is arguably the most well-known analytical model in the framework of the standard flare scenario depicted in 2D. It consists of a pre-existing force-free magnetic flux rope (and its mirror current below the photosphere) in the solar corona. The background coronal magnetic field is created by having a pair of magnetic sources with opposite polarities located at (or slightly below) the photosphere. As shown by ref\cite{1995ApJ...446..377F}, the flux rope can lose its equilibrium due to converging motions of these two foopoint sources and rise, leading to a ``catastrophic'' eruption. The flux rope eruption induces an extended current sheet trailing the rope in which fast magnetic reconnection can be triggered, which further facilitates the eruption through the release of the magnetic energy.

Here we use the formulae of the magnetic vector potential distribution $A(x,y)$ described in ref\cite{2000JGR...105.2375L} to build the analytical magnetic model using observation-constrained free parameters, which only include the height of the flux rope center $h$ (from EUV imaging of the flux rope cavity), footpoint separation $2\lambda$ (from the size of flare arcade at the surface), the location of the lower and upper tip of the RCS $p$ and $q$ (obtained respectively from the tip of the cusp-shaped flare arcades and the bottom of the balloon-shaped flux rope cavity) at different times of the flare event, and a scaling factor $A_0$ for the strength of the photospheric magnetic sources. Examples of the magnetic model overlaid on EUV images at three selected times are shown in the first row of Extended Data Fig. \ref{sfig:model}. An excellent match is found for the flare geometry between the model and the observations. Moreover, after adjusting the scaling factor $A_0$ to match the magnetic field strength according to the values derived from EOVSA microwave data, the coronal magnetic field profile in the close vicinity of the RCS $B(y)$ from the model agrees very well with the measurements from the microwave spectral imaging data (blue curve in Fig. \ref{fig:rec_e}(b)). 

\smallskip

\noindent \textbf{MHD Simulation} \\
We perform self-consistent, 2.5D resistive MHD numerical simulation for this event based on very similar initial setups and scaling in the analytical model described above. The physical parameters in the simulation are homogeneous along the third dimension. Extended Data Fig. \ref{sfig:model} shows the initial setup nearly identical to the analytical model. At this point, the flux rope has risen to a location with a single reconnection $X$ point formed between the rope and the underlying closed arcades (i.e., the initial length of the vertical RCS is zero). The initial height of the flux rope $h$ is adjusted according to the theoretical model in order to place the rope in a state of non-equilibrium for its subsequent eruption \cite{1995ApJ...446..377F, 2000JGR...105.2375L}. Since the evolution starts in a non-equilibrium state, the flux-rope can rise at the beginning with a quick acceleration followed by the formation of an extending RCS at later times.

\renewcommand\thefigure{\arabic{figure}}
\renewcommand{\figurename}{Extended Data Fig.}
\setcounter{figure}{0} 
\begin{figure*}[!ht]
\begin{center}
\includegraphics[width=0.75\textwidth]{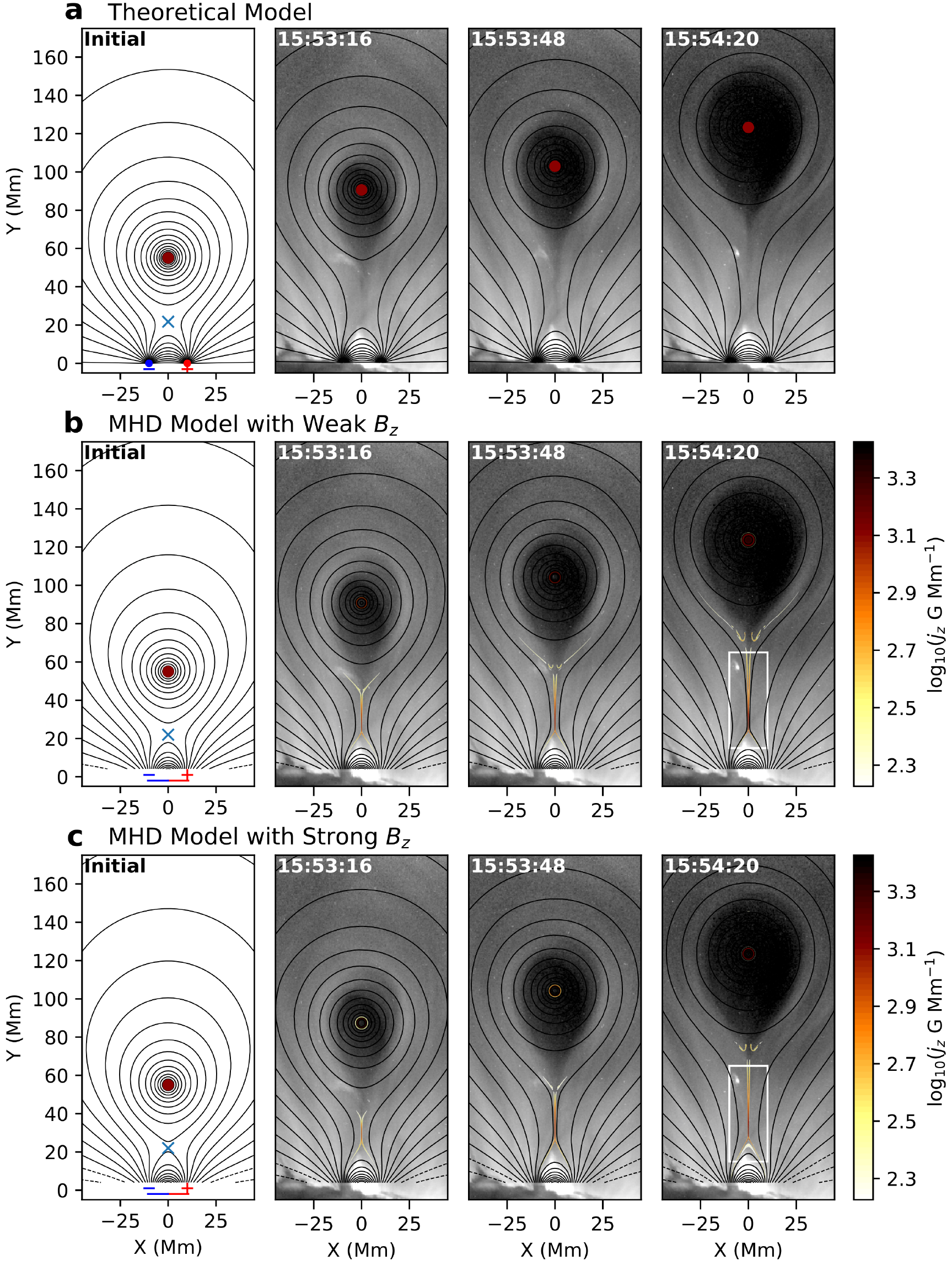}
\end{center}
\caption{\textbf{Magnetic modeling of the X8.2 eruptive solar flare event on 2017 September 10.} (\textbf{a}) Representative magnetic field lines from the analytical standard flare model of Lin \& Forbes\cite{2000JGR...105.2375L}. (\textbf{b} and \textbf{c}) Results from the numerical resistive 2.5D MHD model in the weak and strong guide field $B_z$ case, respectively. Background is SDO/AIA time-series images of the EUV 211 \AA\ filter band. The thin vertical structure with red-orange color near $x=0$ Mm is the reconnection current sheet with an enhanced electric current density $j_z$. The first panel in each row shows the initial conditions of the magnetic modeling, which consist of a line current that represents the magnetic flux rope (red circle symbol) and a pair of bipolar magnetic sources at the solar surface (point sources in theoretical model and line sources in MHD).}
\label{sfig:model}
\end{figure*}

Our simulation box has a grid size of 512 $\times$ 1536. Three levels of adaptive mesh refinement (AMR) are introduced in regions with a large pressure gradient. The finest grid size and typical time step are $2.44 \times 10^{-4}$ and $\sim$10$^{-5}$ in normalized units, which correspond to, respectively, $0.0732$ Mm and ${0.00138}$ s in physical units. The simulation was performed using a publicly available MHD code \textit{Athena++}\cite{2008ApJS..178..137S}, where the hyperbolic MHD parts are solved by the Godunov-type method and shock structures are captured using the Harten-Lax-van Leer-Discontinuities (HLLD) Riemann solver.

\begin{figure*}[!ht]
\begin{center}
\includegraphics[width=1.0\textwidth]{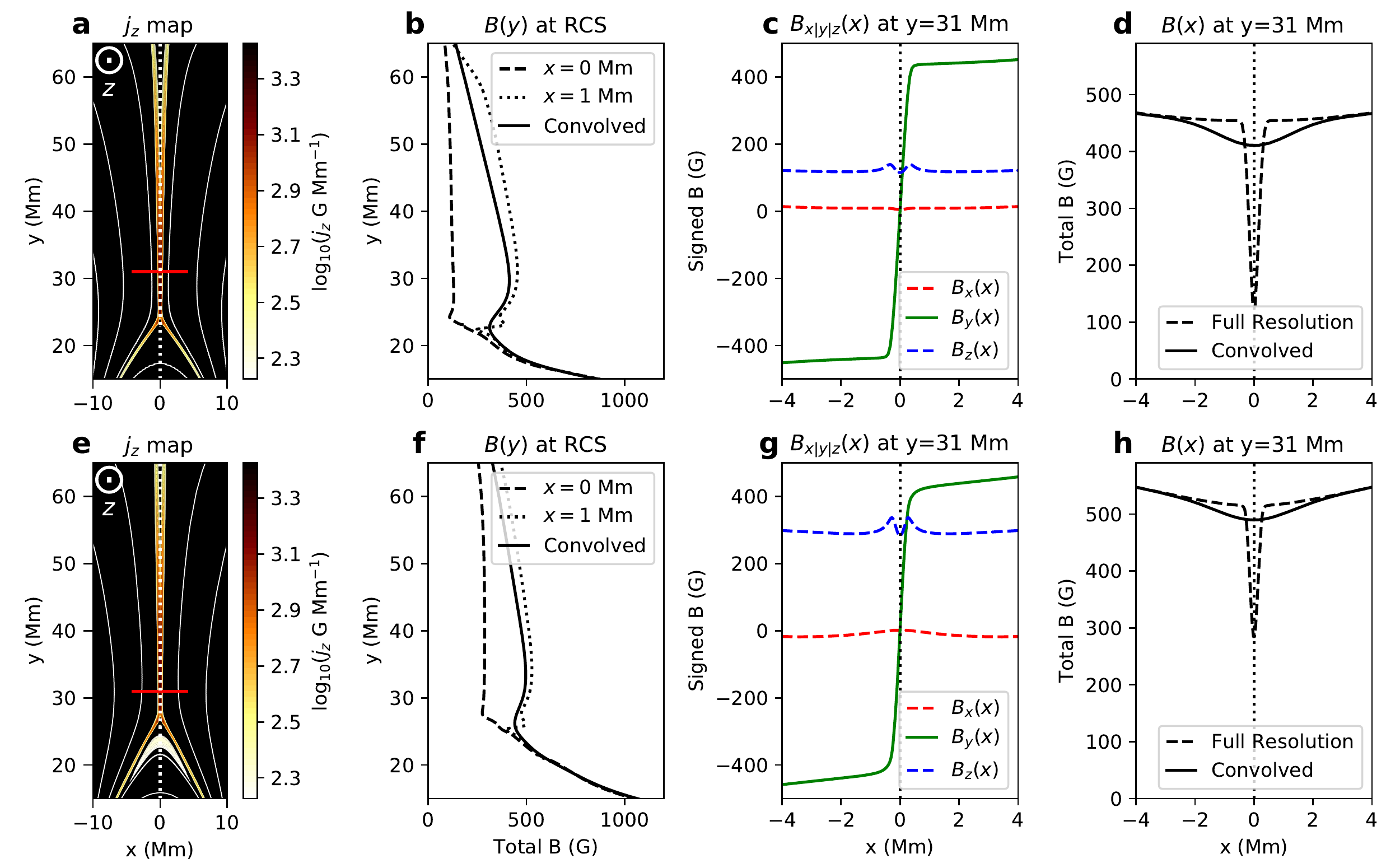}
\end{center}
\caption{\textbf{Magnetic field variation across and along the reconnection current sheet in MHD simulation.} (\textbf{a}) Enlarged view of the central RCS region in the MHD model (white box in the right panels of Extended Data Fig. \ref{sfig:model}(b) and (c)). The RCS exhibits itself as the vertical feature with a strong current density $j_z$. (\textbf{b}) Height variation of the total magnetic field strength $B(y)$ along the RCS (at $x = 0$ Mm; vertical dashed line in (a)). Dashed and dotted curves represent results from the full-resolution MHD model at $x=0$ Mm and $x=1$ Mm. Solid curve is the $B(y)$ profile obtained after convolution with EOVSA's instrument resolution. The latter contains key information about the average magnetic field in the immediate vicinity of the RCS (same as the red curve in Fig. \ref{fig:rec_e}(b)), which compares favorably with results derived from EOVSA microwave observations. (\textbf{c}) Spatial variation of the $x$, $y$, $z$ components of the magnetic field vector across the RCS ($B_x(x)$, $B_y(x)$, $B_z(x)$) obtained at $y = 31$ Mm (horizontal line in (a)). (\textbf{d}) Total magnetic field variation across the RCS $B(x)$. Dashed and solid curves show the result from the full-resolution MHD model and that after convolution with EOVSA instrument resolution. Note the sharp dip at the very center of the current sheet is smoothed out. (\textbf{e})--(\textbf{h}) Same as above, but for the stronger guide field $B_z$ case.} \label{sfig:mhd_cs}
\end{figure*}

One notable modification from the analytical model lies in the magnetic sources at the bottom boundary: the two point sources at the photosphere in the original theoretical model are replaced with a pair of extended line sources, shown as blue and red lines in the bottom left panel of Extended Data Fig. \ref{sfig:model}. The reason of such a modification is twofold: First, it is more realistic in the sense that the opposite polarities of the sunspot group in the active region are not point-like, but both show a substantial spatial extension ($>$10 Mm) and are separated by a well-defined polarity inversion line (see, e.g., studies on the photospheric magnetic field of the active region measured a few days before \cite{2018ApJ...868..148L,2018ApJ...869...13J,2018RNAAS...2a...8W}). Second, the difficulty in numerically modeling the area close to the two delta-function foot-point sources is removed. The magnetic field outside the flux-rope is similar to the previous works based on the theoretical model \cite{1990SoPh..126..319P}, except that we introduce a weak current density $j_z$ distributed around the flux-rope (with a Gaussian shape; amounts to $\sim$0.05\% of the maximum current density of the flux-rope) to smooth the sharp edge around the flux-rope. In order to achieve pressure balance and an initial force-free condition within the flux rope, we also introduce a guide field ${B_z}$ (i.e., along the 3rd dimension perpendicular to the $x$-$y$ plane) which peaks at the flux rope center but decreases rapidly at greater distance from the rope. A similar setup of the ${j_z}$ and ${B_z}$ distribution of the flux rope was used in \cite{2019MNRAS.482..588Y}. Lastly, for the purpose of simplification, the coronal background is initialized with a uniform plasma density of $\sim {10^9}$ cm${^{-3}}$ in most of the simulation domain (which only increases toward the flux rope center for the purpose of pressure balance) and a temperature of 2 MK. To facilitate fast magnetic reconnection, we also include a considerable resistivity that corresponds to a magnetic Reynolds number of the order ${R_m \sim 10^5}$. Different selections of the $R_m$ value would affect the internal properties of the RCS and flare dynamics. However, it has little impact on the large-scale magnetic configuration surrounding the RCS, which is the primary focus of this modeling study.

In the 2.5-D MHD model, the RCS exhibits itself as a vertical feature with a strong current density $j_z$ located at $x=0$ Mm (Extended Data Fig. \ref{sfig:mhd_cs}(a)). Extended Data Fig. \ref{sfig:mhd_cs}(c) demonstrates the variation of the $x$, $y$, $z$ components of the magnetic field vector across the current sheet (i.e., $B_x(x)$, $B_y(x)$, $B_z(x)$) at a selected height close to the reconnection $X$ point. At the center of the RCS, the vertical component $B_y$ quickly switches its sign and the horizontal component $B_x$ is nearly zero. Both phenomena are characteristics of ongoing magnetic reconnection in the RCS, which is responsible for releasing the magnetic energy and powering the flare \cite{priest_forbes_2000}. The total magnetic field strength $B = \sqrt{B_x^2 + B_y^2 + B_z^2}$ shown in Extended Data Fig. \ref{sfig:mhd_cs}(d), therefore, displays a very sharp and narrow ($<$400 km in width) dip at the RCS center. For comparing the MHD modeling results directly with the magnetic field measurements from observations with finite resolution, we have convolved the magnetic field distribution in the MHD model using a Gaussian function with a full-width-half-maximum of 3 Mm (equivalent to EOVSA's resolution at $\nu=15$ GHz). After the convolution, the sharp dip in total magnetic field across the RCS is nearly smoothed out (solid curve in Extended Data Fig. \ref{sfig:mhd_cs}(d)). However, the spatial variation of the total magnetic field as a function of height ($B(y)$) in the immediate vicinity of the RCS is preserved (Extended Data Fig. \ref{sfig:mhd_cs}(b)), which, as we discussed in the main text, allows us to identify the reconnection $X$ point as a local maximum and the looptop ``magnetic bottle'' as a local minimum on the $B(y)$ profile.

The perpendicular component of the magnetic field $B_z\approx B\cos\theta$, usually referred to as the ``guide field'', may have a profound impact on the detailed reconnection and particle acceleration processes \cite{2009JPlPh..75..159Z, 2017PhPl...24i2110D}. To investigate the possible impacts of the presence of a guide field $B_z$ on the overall flare geometry, we have run two MHD test cases. The first case has a relatively weak $B_z$, which amounts to $\sim$30\% of the total magnetic field strength $B$ in the RCS region (corresponding to a viewing angle $\theta=\arccos{(B_z/B)}\approx 70^{\circ}$, a typical value derived from the microwave spectral fit results). In the second case, a stronger guide field of $\sim$60\% of the total field strength is introduced (corresponding to $\theta\approx 50^{\circ}$, which is near the lower-bound of typical fit values). The results of the overall magnetic geometry for the two cases are shown in Extended Data Fig. \ref{sfig:model}(b) and (c), respectively. More detailed variations of the magnetic field components in the RCS region for the two cases are shown in Extended Data Fig. \ref{sfig:mhd_cs} (top and bottom row). We find that, although the dynamics of the magnetic flux rope eruption differ slightly between the two cases, the overall flare geometry exhibits very little differences. However, detailed features of the magnetic field strength profile at the RCS $B(y)$, including the local maximum and minimum near the reconnection $X$ and $Y$ point, are affected by the different values of the guide field introduced in the MHD model---e.g., a strong $B_z$ throughout the simulation domain would make the peculiar features associated with the reconnection current sheet less profound (see, e.g., the comparison between Extended Data Fig. \ref{sfig:mhd_cs}(b) and (f)). In this work, we find a better match of the $B(y)$ profile between our observations and the weak guide field case, which we adopt in the observation--modeling comparison.

Our self-consistent 2.5D modeling matches the observed flare geometry and RCS magnetic field profile as the theoretical magnetic model (Extended Data Fig. \ref{sfig:model}(b) and Figure \ref{fig:rec_e}(b)). It also provides a crucial framework for us to identify various key components associated with the magnetic reconnection, which include the RCS and the primary reconnection X point, the plasma inflows and outflows, and the distribution of the reconnecting magnetic energy and electric field along the RCS.

\subsection*{Microwave Spectral Analysis}
The EOVSA instrument and an overview of the observation of the 2017 September 10 X8.2 flare were discussed in a recent paper\cite{2018ApJ...863...83G}. To briefly summarize, EOVSA obtained data in 2.5--18 GHz of this event with 134 frequency channels spread over 31 equally spaced spectral windows (SPWs), each of which has a bandwidth of 160 MHz. The center frequencies of these SPWs are given by $\nu=2.92 + n/2$ GHz, where $n$ is the SPW number from 0 to 30. In this study, images were made in 3.4--18 GHz by combining the spectral channels within each of SPWs 1--30 using the CLEAN algorithm. The nominal full-width-half-max (FWHM) angular resolution is $113''.7/\nu_{\rm GHz}\times53''.0/\nu_{\rm GHz}$ at the time of the observation. In this study, a circular beam with a FWHM size of $73''.0/\nu_{\rm GHz}$ is used for restoring the CLEAN images, while the size is fixed at 5$''$ above 14.5 GHz (note that here we used a slightly smaller restoring beam than ref\cite{2018ApJ...863...83G}). 

Microwave spectral imaging data from EOVSA allow us to derive a microwave spectrum $F(\nu)$ at each selected pixel location ($x$ and $y$) and time $t$. The spatially- and temporally-resolved microwave spectra show characteristics of the gyrosynchrotron radiation produced by energetic electrons gyrating in the coronal magnetic field\cite{1982ApJ...259..350D}. Here we employ the fast gyrosynchrotron codes\cite{2010ApJ...721.1127F} to calculate the microwave brightness temperature spectra based on the gyrosynchrotron radiation theory. The codes perform full radiative transfer calculation along the line of sight (LOS; approximately the $z$ direction in our adopted coordinate system, with $x$- and $y$-axes aligned with solar south-north and east-west, respectively), with the capability of reducing the computing time by several orders of magnitude compared with approaches that use exact formulae\cite{1969ApJ...158..753R} while retaining the accuracy of the resulting spectra. 

The spatially-resolved microwave spectra contain information about the flare-accelerated energetic electrons, particularly those at mildly relativistic energies, as well as unique diagnostics for the magnetic field strength in the source region. The peak frequency of the spectra is sensitive to the magnetic field strength $B$ and the number density of energetic electrons $n_e$. The high-frequency, optically-thin side of the spectra is mainly determined by the electron energy distribution with a spectral index $\delta$. The low-frequency, optically-thick side of the spectra constrains the effective temperature of the nonthermal electrons and to some extent, density and temperature of thermal plasma if free-free absorption or Razin suppression play a  role. For more details on the diagnostics of the source parameters using microwave gyrosynchrotron spectra, we refer the readers to other works.\cite{2010ApJ...721.1127F, 2004ASSL..314...71G,2009ApJ...698L.183F,2013SoPh..288..549G,2018ApJ...863...83G,Fleishman2020} Although the gyrosynchrotron radiation spectra have the potential to constrain flare-accelerated nonthermal electrons in a broad range of energies from a dozen keV to MeV range, for this study, we focus on those at mildly relativistic energies ($\sim$100 keV--1 MeV).

\begin{figure*}[!ht]
\begin{center}
\includegraphics[width=0.9\textwidth]{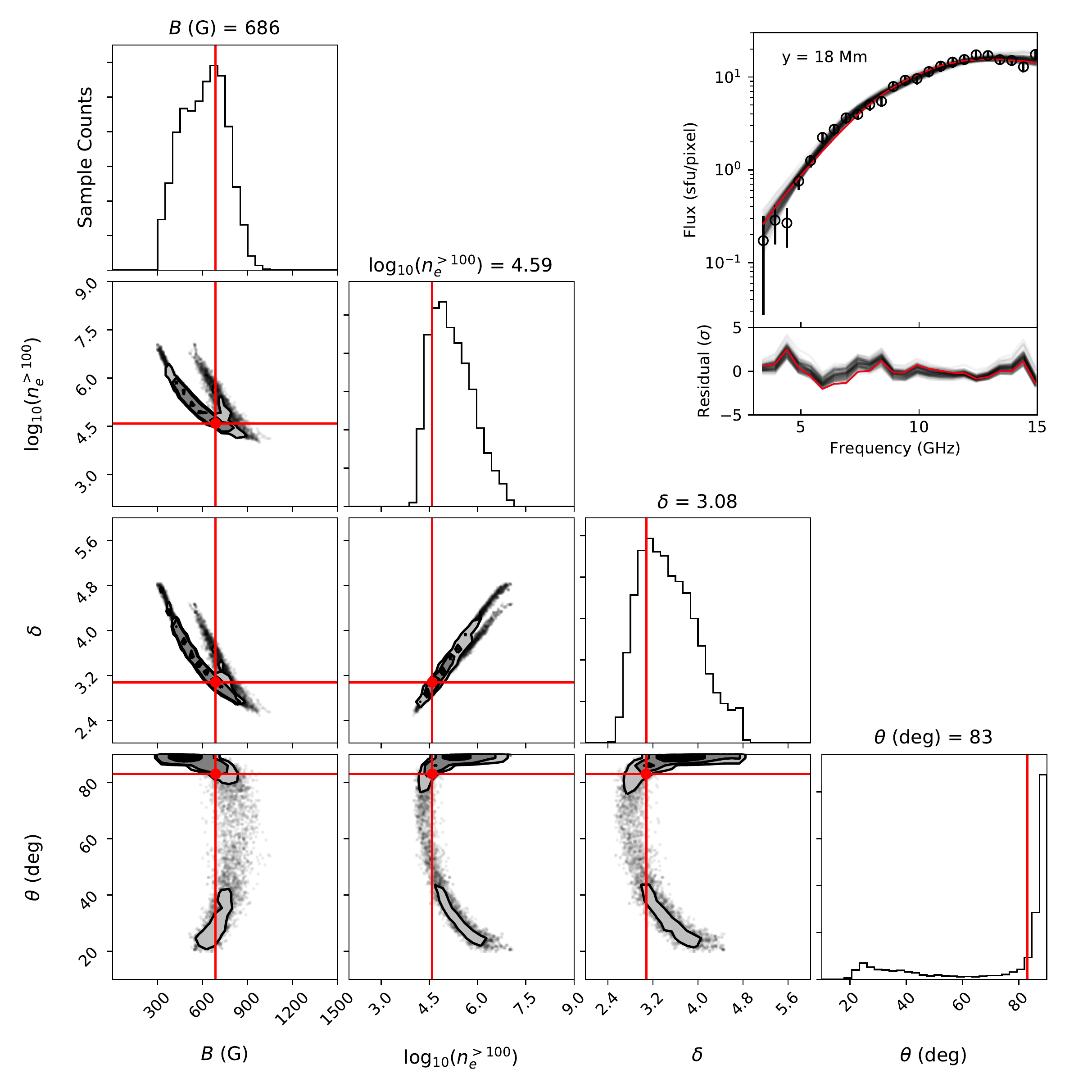}
\end{center}
\caption{\textbf{Markov chain Monte Carlo analysis for an example spatially resolved microwave spectrum}. The spectrum is taken from the location labeled ``2'' in Fig. \ref{fig:spec}(c). Red lines/circles in each panel indicate the final fit results from the MCMC analysis. Corresponding spectra and residuals calculated from each MCMC sampling in the multi-parameter space are shown in the upper right panel as gray curves. Red curves are the final fit spectrum and residual. Note the total number density of energetic electrons shown in the corner plot is the result integrated above 100 keV ($n^{>100}_e$), which is different from the value of $n^{>300}_e$ shown in Fig. \ref{fig:spec}(d).} \label{sfig:mcmc_1src}
\end{figure*}

\begin{figure*}[!ht]
\begin{center}
\includegraphics[width=1.0\textwidth]{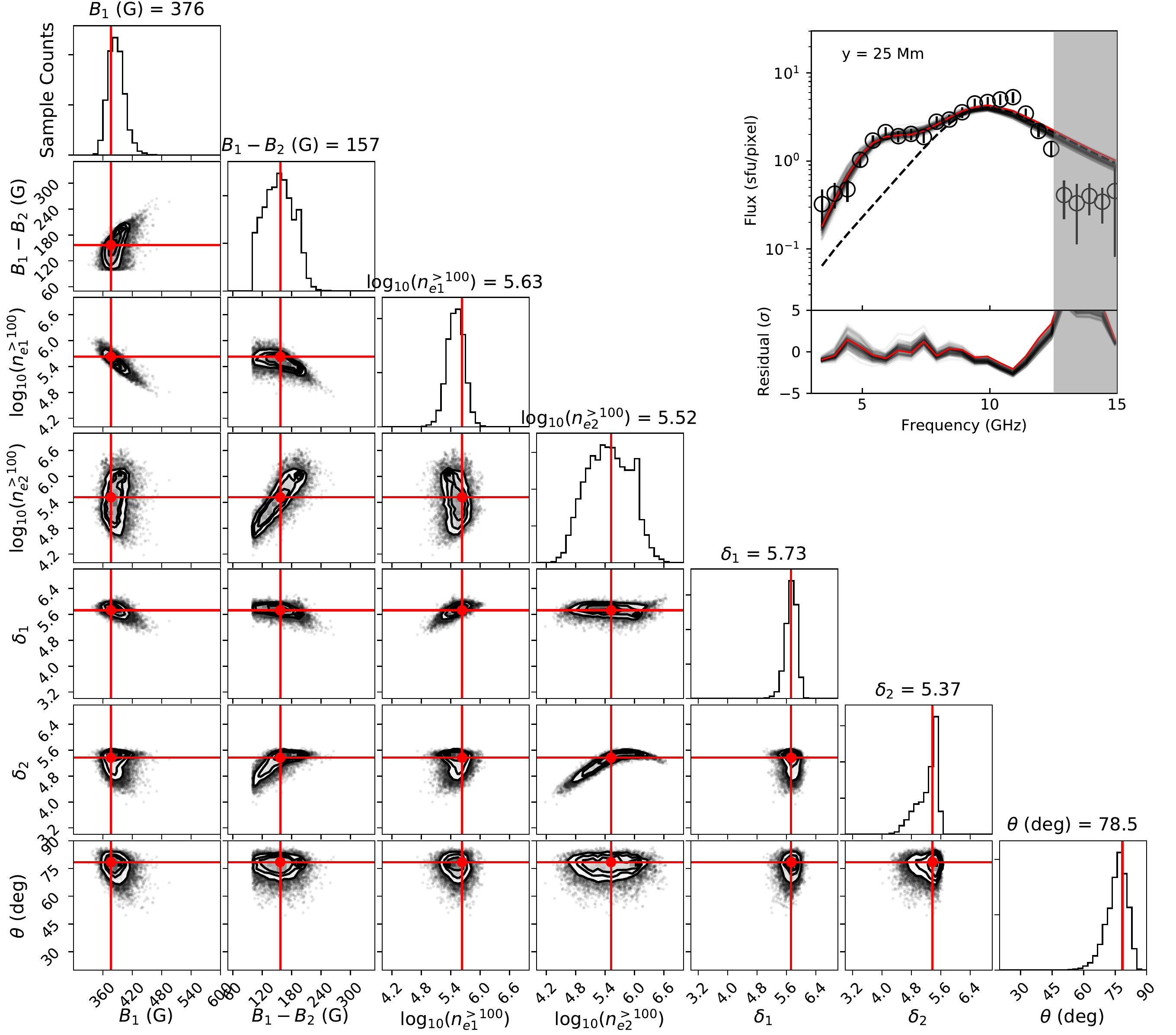}
\end{center}
\caption{\textbf{Markov chain Monte Carlo analysis for an example spatially resolved microwave spectrum with two spectral components}. The spectrum is taken from the location labeled ``3'' in Fig. \ref{fig:spec}(c). The corner plots are similar to Extended Data Fig. \ref{sfig:mcmc_1src}, but they show MCMC results with two source components. Parameters with subscripts ``1'' and ``2'' indicate the physical parameter for the two components, respectively. Red curve and dashed black curve in the upper right panel shows, respectively, the fit spectrum with both components and the spectrum calculated from the component with a stronger magnetic field only (i.e., component with subscript ``1'').} \label{sfig:mcmc_2src}
\end{figure*}

Here we adopt an algorithm\cite{2009ApJ...698L.183F,Fleishman2020} to fit the spatially-resolved microwave spectra to obtain an initial set of physical parameters of the source, which include the magnetic field strength $B$, the angle between the magnetic field vector and the LOS direction $\theta$, the energetic electron distribution $f_e(\epsilon)$, and the thermal electron density $n^{th}_e$. We assume a homogeneous source along the LOS with a column depth of 10$''$, as well as a power-law electron energy distribution $f_e(\varepsilon)$ with a spectral index $\delta$, and low- and high-energy cutoff of 10 keV and 10 MeV, respectively. As already verified by detailed tests using simulated microwave spectra from realistic 3D flare models \cite{2009ApJ...698L.183F,2013SoPh..288..549G}, the fit algorithm works very well to recover the source parameters for spectra with a single, well-defined peak located within the observational frequency range. However, there are a few cases that pose challenges for the algorithm: (1) For spectra at lower heights where the magnetic field strength is particularly high, the spectra appear to continue to rise beyond the highest observable frequency (e.g., bottom right panel of Fig. \ref{fig:spec}(d)), such that the spectral peak is absent. (2) For spectra at higher heights, the high-frequency portion of the spectra is largely dominated by noise and could not be included for spectral analysis (shadowed area in Fig. \ref{fig:spec}(d)). This is largely limited by the signal-to-noise-ratio (SNR) of the instrument (up to $\sim$100 for EOVSA): the presence of a very bright high-frequency source at lower heights (in the looptop region) hinders the detectability for a much weaker source at greater heights. (3) At some other locations, the spectra display more than one spectral peak, which implies the presence of multiple components within the resolution element. 

In order to evaluate and refine the initial fit results, we employ a Markov chain Monte Carlo (MCMC) analysis method, implemented by an open-source Python package \textit{emcee} \cite{2013PASP..125..306F}, to sample the posterior probability distributions (PPDs) of the fit results based on Bayesian statistics \cite{2010CAMCS...5...65G}. We have performed such MCMC analysis for all the microwave spectra along the current sheet. Extended Data Fig. \ref{sfig:mcmc_1src} shows an example of the MCMC analysis results in the form of a ``corner plot''. In the corner plot, the diagonal panels show the one-dimensional projection of the PPDs of the respective fit parameters. The two-dimensional projections of the PPDs between pairs of the fit parameters are shown as the non-diagonal panels. These probability distributions provide quantitative constraints on the most probable locations to find the fit parameters in the multi-parameter space. The widths of the PPDs are, in turn, optimal estimates for the uncertainties of the respective fit parameters. As expected, for a spectrum that has a single spectral peak in the observing frequency range, the PPDs of the fit parameters are clustered around the minimization results, such that the fit results are well constrained. If the spectral peak is not very profound or is completely absent from the observing frequency range, the PPDs are relatively broader, and sometimes display more than one local concentration in the multi-parameter PPDs. For spectra at higher heights with noisy measurements at high frequencies, the broader PPDs are also present. For these cases, the fit results of the respective parameters have larger uncertainties and, under some circumstances, are not unique. The increased uncertainties for these spatial locations are reflected by the larger error bars shown in Fig. \ref{fig:rec_e}. For these cases, we use fit results from nearby pixels (with well constrained spectra) to inform the selection of the appropriate range of the fit parameters. Another round of spectral fit is then performed to ensure that the resulting fit parameters conform with the PPDs from the MCMC analysis. Extended Data Fig. \ref{sfig:mcmc_1src} shows an example of a marginal case in which the spectral peak is not very profound (which correspond to location ``2'' in Fig. \ref{fig:spec}(c).
Although the multi-parameter PPDs display more than one branches of distribution, the MCMC approach successfully finds the most probable combination of parameters that also achieves a good fit of the observed spectrum. We caution that, however, the best multi-parameter fit results do not necessarily always coincide with the peak value(s) in a given 1D or 2D PPD in the corner plot for a given parameter or parameter pair.

At a small subset of spatial locations (at $y\approx21$--28 Mm around the above-the-loop-top region near the bottom of the RCS), the spectra display a secondary spectral peak. This is possible indication for the existence of a second population of accelerated electrons in this highly dynamic region where reconnection outflows meet the newly reconnected flare arcade. Such spectra could not be fit with a model that only assumes one homogeneous source along the LOS. For these cases, we introduce a secondary source along the LOS that shares the same parameters as the primary source but differs only in $B$, $n_e$, and $\delta$. The fit results are again evaluated using the MCMC method, and the associated uncertainties are reported accordingly. As demonstrated in Extended Data Fig. \ref{sfig:mcmc_2src}, although the degree of freedom is inevitably increased with the addition of more fit parameters, there are adequate measured data points in the microwave spectra to warrant a reliable fit as evidenced by the well-defined PPDs of the fit parameters. For these spectra, we show the resulting magnetic field $B$ associated with the primary component (with a higher $B$ value) in Fig. \ref{fig:rec_e}(b), and the total $n_e$ values from both components in Figs. \ref{fig:spec}(d) and \ref{fig:rec_e}(e).

We note that the coronal magnetic field strength derived from the microwave data is consistent with the results from ref\cite{2019ApJ...874..126K}, who reported a coronal field strength of up to 350 G at a height of $\sim$25 Mm in the post-flare arcade using infrared spectropolarimetry based on measurements of the magnetically sensitive Ca II 8542 \AA\ line. Our measurements of a strong coronal magnetic field is also consistent with the measurements of multi-kilogauss (up to $>$5000 G) photospheric field in the core region of the active region when the same region was viewed on disk four days before \cite{2018RNAAS...2a...8W}, as well as the coronal magnetic field extrapolated from the photospheric measurements and validated using high-frequency microwave probing of the coronal magnetic field \cite{Anfinogentov_2019}. 

\subsection*{EUV Plasma Flows}
To investigate plasma flows in the close vicinity of the magnetic reconnection site and measure their speeds in the plane of the sky, we use observations from the Atmospheric Imaging Assembly aboard the Solar Dynamics Observatory (SDO/AIA)\cite{2012SoPh..275...17L}, which provides full-Sun imaging at multiple EUV filter bands with a spatial resolution of $\sim$1.2$''$ (pixel size $0''.6$) and a cadence of 12~s. To reveal plasma flows along the direction of the RCS, we make a vertical slice at a location along RCS (labelled slice ``a'' in Fig. \ref{fig:flows}(b)). At each spatial location at the slice $y$, we obtain the time evolution of the EUV intensity $I(t,y)$, which is displayed in the form of a ``time-distance plot'', shown in Fig. \ref{fig:flows}(c). In the time-distance plot, the horizontal and vertical axes represent time and spatial location along the slice, respectively. We also apply a running-ratio technique on the time-distance plots in order to bring out the fast time-varying features (i.e., plasma flows): the normalized intensity shown at each time and spatial pixel ($(t, y)$) is the ratio of the original intensity $I(t,y)$ to its second nearest neighbor frame at the same location $y$. The same technique is applied to all SDO/AIA EUV passband images. We find that, at the time of interest, the plasma flows along the direction of the RCS (i.e., the vertical direction $y$ near $x\approx 0$ Mm) are best seen in the 171 \AA\ and 211 \AA\ passbands, possibly due to their sensitivity to continuum emission (thermal bremsstrahlung) at flare temperatures \cite{2010A&A...521A..21O,2018ApJ...854..122W}. In the SDO/AIA 171 \AA time-distance plot of Fig. \ref{fig:flows}(c), downward-moving plasma flows appear just below the bottom of the RCS (or the reconnection $Y$ point; at $y\approx 21$ Mm) as coherent tracks moving toward the bottom-right direction. In the accompanying SDO/AIA 171 \AA\ running-ratio animation (Supplementary Video 1), we find that these downward-moving plasma flows are associated with the fast contraction motion of the newly reconnected loops emanating from the tip of the cusp-shaped feature (located near the RCS bottom). The speeds of the contracting loops are measured using the slopes of these tracks in the time-distance plot, which amount to $\sim$150--510 km s$^{-1}$. 

The observed speeds of the plasma downflows (or fast-contracting loops) below the bottom of the RCS ($\sim$150--510 km s$^{-1}$) are at least an order of magnitude slower than the estimated Alfv\'en speeds in the inflow region, $\sim$6,000--10,000 km s$^{-1}$. This result is in line with previous findings on plasma flows above the post-flare arcades: it have been shown that virtually all reported signatures of plasma outflows, including the so-called supra-arcade downflows (SADs) and supra-arcade downflowing loops (SADLs), have velocities well below the presumed reconnection outflows at or close to Alfv\'en speeds \cite{2010ApJ...722..329S,2011ApJ...730...98S,2018ApJ...868..148L}. Such a persistent speed discrepancy has been discussed in the literature (see discussions in \cite{2018ApJ...868..148L} and references therein). Here we highlight one possibility: high-speed Alfv\'enic plasma outflows are too fast to be detected in EUV/SXR time-series images with a limited time cadence---in this case, outflows at Alfv\'en speeds would traverse the entire length of the RCS ($\sim$50 Mm at the time of interest) within $\sim$0.5 s, much shorter than AIA's cadence of 12 s. In order to readily detect these Alfv\'enic plasma flows through running-difference/ratio imaging based on a few neighboring time integrations, the flows need to be slowed down substantially to $\lesssim$1,000-2,000 km s$^{-1}$ (as in our case and many other reported cases in the literature) due to, e.g., a drag force along its path.

To investigate plasma inflows at different locations of the RCS $v_x(y)$, we make a series of horizontal slices across the RCS at different heights (labeled ``b1'' to ``b5'' in Fig. \ref{fig:flows}(b)). For each slice at a height $y$, we obtain the EUV intensity at all the pixels on the slice (i.e., in the $x$ direction) as a function of time, resulting in a series of time-distance plots shown in Fig. \ref{fig:flows}(a). The plasma inflows appear as close-to-linear tracks on the running-ratio time-distance plots, whose speeds are measured based on their slopes. The uncertainties of the inflow speed measurements are estimated empirically by assuming a spatial uncertainty of four AIA pixels (2.4$''$, or about $2 \times$ AIA angular resolution) for each position measurement, together with a temporal uncertainty of 12 s (i.e., $1\times$ AIA cadence) for the time determination. We note that, as shown from the time-distance plots in Fig. \ref{fig:flows}(a) and the accompanying animation (Supplementary Video 1), the converging inflows seem to evolve slightly toward the $-x$ direction at later times. This is likely due to the temporal evolution of the current sheet as the flare reconnection progresses.

\subsection*{Powering the Second Largest Solar Flare of Solar Cycle 24}
From measurements of the reconnecting magnetic field $B$ and inflowing plasma speed $v_x$, we obtain an electromagnetic energy flux brought into the RCS for reconnection $S_{\rm rec}$ is of order $10^{10}$--$10^{11}$ ergs s$^{-1}$ cm$^{-2}$. The total energy available for release during the flare impulsive phase is $\dot{\varepsilon}_{\rm rec}=S_{\rm rec}A$, where $A=2l_yl_z$ is the total area of the RCS that is currently undergoing fast reconnection. The length of the RCS $l_y$ is readily available from the microwave/EUV imaging data ($\sim$40 Mm; c.f., Fig. \ref{fig:rec_e}(a)). The depth of the RCS $l_z$ is unknown since it lies along the LOS direction. We take it to be as the same order of the RCS length, 10 Mm. Thus $\dot{\varepsilon}_{\rm rec}\approx10^{29}$--$10^{30}$ ergs s$^{-1}$. As stated in the main text, this is sufficient to power a large X-class flare that releases 10$^{32}$ ergs in several minutes at its peak rate.

\subsection*{Supplementary Video}
\textbf{Supplementary Video 1 \textbar\ Animation accompanying Fig. \ref{fig:flows}}. The animation shows the flare evolution from 15:51:45 UT to 16:06:09 UT on 2017 September 10. Panels (a) and (c) are identical to those in Fig. 4. Panel (b) shows SDO/AIA 171 \AA\ running-ratio time-series images. Examples of the plasma inflows converging toward the RCS from the $-x$ and $+x$ sides are marked in the $x$-$t$ plot in (a) as blue and red curves and in (b) as triangles with the same color. Plasma downflows below the RCS are marked as green curves in the $t$-$y$ plot in (c) and in (b) as green triangles. The moving horizontal/vertical bar in panels (a)/(c) indicates the corresponding time. The animaiton can be accessed at \href{http://harp.njit.edu/~binchen/download/publications/Chen+2020_RCS/Chen_Supplementary_Video_1.mp4}{this URL link}.



\section*{Acknowledgments}
EOVSA operation is supported by NSF grant AST-1910354. The work is supported partly by NASA DRIVE Science Center grant 80NSSC20K0627. B.C., D.G., G.F., G.N., and S.Y. are supported by NASA grants 80NSSC18K1128, 80NSSC19K0068, and NSF grants AGS-1654382, AGS-1723436, and AST-1735405 to NJIT. K.R. and C.S. are supported by NASA grant NNX17AB82G and NSF grants AGS-1723425, AGS-1723313 and AST-1735525 to SAO. F.G. is supported by NSF grant AST-1735414 and DOE grant DE-SC0018240. S.K. is supported by NASA contract NAS 5-98033 for RHESSI. J.L. is supported by the Strategic Priority Research Program of CAS with grants XDA17040507, QYZDJ-SSWSLH012, XDA15010900, NSFC grants U1631130, the project of the Group for Innovation of Yunnan Province grant 2018HC023, and the Yunnan Yunling Scholar Project. X.K. is supported by the NSFC grants 11873036 and 11790303 (11790300), the Young Elite Scientists Sponsorship Program by CAST, and the Young Scholars Program of Shandong University. The MHD simulations performed for this work were conducted on the Smithsonian High Performance Cluster of Smithsonian Institution, and used resources of the National Energy Research Scientific Computing Center. This work made use of public software packages CASA, SunPy, Astropy, Athena++, emcee, and lmfit.

\section*{Author Contributions}
B.C. conceived the study, carried out the data reduction, analysis, interpretation, and manuscript preparation. C.S. performed the MHD simulation and contributed to observation--modeling comparison. D.G. led the construction and operation of EOVSA, and contributed to microwave data calibration and interpretation. K.R. provided codes for the theoretical magnetic model and contributed to the observation--modeling comparison. G.F. provided codes for calculating gyrosynchrotron radiation and contributed to microwave spectral fitting. S.Y. contributed to microwave data calibration and EUV data analysis. S.K. performed HXR imaging and contributed to the interpretation of data. J.L. contributed to MHD simulation. F.G. and X.K. contributed to the interpretation of data. G.N. contributed to microwave spectral fitting. All authors discussed the results and contributed to manuscript preparation.


\end{document}